\documentstyle[twoside,fleqn,espcrc2,epsf]{article}

\newcommand{\ttbs}{\char'134}
\newcommand{\AmS}{{\protect\the\textfont2
  A\kern-.1667em\lower.5ex\hbox{M}\kern-.125emS}}

\def\partialslash{\not{\hskip -0.1cm}\partial}
\newcommand{\msbar}{\overline{\rm{MS}}}

\newcommand{\pslash}{p \!\!\!/}
\newcommand{\xslash}{x \!\!\!/}
\newcommand{\Dslash}{D \!\!\!/}
\newcommand{\nn}{\nonumber}
\newcommand{\beq}{\begin{eqnarray}}
\newcommand{\eeq}{\end{eqnarray}}
\newcommand{\eq}{\ref}
\newcommand{\np}{Nucl.Phys.\ }
\newcommand{\pl}{Phys.Lett.\ }
\newcommand{\pr}{Phys.Rev.\ }
\newcommand{\asgen}{\alpha_s}
\newcommand{\asl}{\alpha_s^{{}^{LATT}}}
\newcommand{\as}{\alpha_s^{{}^{\widetilde{MOM}}}}
\newcommand{\asms}{\alpha_{{}^{\overline{MS}}}}
\newcommand{\MSB}{\overline{MS}}
\newcommand{\MOM}{\widetilde{MOM}}
\newcommand{\gduems}{g^2_{\overline{MS}}}
\newcommand{\gdue}{g^2_{\widetilde{MOM}}}
\newcommand{\Lam}{\Lambda_{\widetilde{MOM}}}
\newcommand{\Lams}{\Lambda_{\overline{MS}}}
\newcommand{\Lamlatt}{\Lambda_{{}^{LATT}}}
\newcommand{\Gev}{{\rm GeV}}
\newcommand{\Mev}{{\rm MeV}}
\newcommand{\liege}{\address{Institut de Physique,
Universit\'e de Li\`ege au Sart Tilman, B-4000 Li\`ege, Belgique.} }    

\newcommand{\lpthe}{\address{L.P.T.H.E., Universit\'e de Paris Sud, 
Centre d'Orsay, 91405 Orsay, France.} }

\newcommand{\tov}{\address{Dipartimento di Fisica, 
Universit\`a di Roma \lq Tor 
Vergata\rq \,\,  and INFN, Sezione di Roma II, \\
~Via della Ricerca Scientifica 1, I-00133 Rome, Italy.} }    

\newcommand{\rom}{\address{Dipartimento di Fisica, 
Universit\`a di Roma \lq La 
Sapienza\rq \,\, and INFN, Sezione di Roma I, \\
~P.le A. Moro, I-00185 Rome, Italy.} }

\newcommand{\cern}{\address{Theory Division, CERN, CH-1211 Geneva 23,
Switzerland.} }

\newcommand{\south}{\address{Department of Physics,    
The University, Southampton SO9 5NH, U.K.} }

\title{Pseudoscalar Vertex and Quark Masses}

\author{Jean-Ren\'e Cudell\liege, 
Alain Le Yaouanc\lpthe
 and 
Carlotta Pittori$^a$%
    \thanks{Talk presented by Carlotta Pittori at LATTICE99}}

\begin{document}

\begin{abstract}We analyse available data on the quark pseudoscalar vertex and
extract the contribution of the Goldstone boson pole. The strength of the
pole is found to be quite large at presently accessible scales. We draw the
important consequences of this finding for the various definitions of quark
masses. 
\end{abstract}

\maketitle

\section{Introduction}
The expected behaviour of the 
quark pseudoscalar (PS) vertex in the continuum,
near the chiral limit, can be described through a
perturbative contribution plus a non-perturbative 
Goldstone boson contribution.
%
%
According to the Wilson operator product expansion (OPE),
the non-perturbative
contribution  must be power behaved, {\it i.e.} at large $p^2$ it must
drop as $1/p^2$ up to logs. 
On the lattice, the use of a non perturbative renormalisation scheme
\cite{rome} makes this contribution manifest.
Although it
goes to zero for large momentum transfers, our purpose 
is to extract it from lattice simulation data, and to show
that it is not negligible for presently
accessible scales.
Moreover, from the axial Ward identity (AWI),
the forward PS vertex is directly
related to the scalar part of the quark propagator.
Hence 
in the study of propagator OPE, 
the Goldstone pole contribution
in the PS vertex should corresponds to 
the dominant non-perturbative (power) contribution in
the quark condensate.
In the following we will show our recent results, presented in ref.\cite{pio},
of the analysis of available
lattice data on the quark pseudoscalar vertex. We have
extracted the Goldstone contribution, in $1/p^2$ and the strength
of the pole is found to be quite large at presently accessible scales.
We draw the important consequences of
this finding for the various definitions of quark masses.
Finally we will present some preliminary results of our work in progress
\cite{prop} on the lattice quark propagator analysis.
\section{The quark PS vertex:
results from the fit to lattice data}
In order to evaluate the non-perturbative Goldstone 
contribution, we have analised the 
QCDSF collaboration data \cite{rakow}, at $\beta=6.0$,
for the bare one-particle-irreducible lattice PS vertex,
$\Lambda_5$,
presented through the product (see their fig.1)
\beq
 am_q~\tilde\Gamma_5(p^2)\equiv C\times ~am_q ~\Gamma_5 (q=0,p^2)
\label{gamma5}
\eeq
at three $\kappa$ values, where
$\Gamma_5\equiv Tr(\gamma_5 \Lambda_5)/4$ and
by construction, $C$ is a constant with the value $C=0.75$
in order to satisfy the AWI in a momentum range, increasing with
$\kappa$.\\
As
$\Gamma_5 (q=0,p^2)\propto \left[A(p^2)/am_q+B(p^2)\right]$,
it seems reasonable to
identify
the $A(p^2)$ term as the
Goldstone contribution and the second one as
the perturbative contribution,
if we are sufficiently close to the chiral limit.
Let us fit the data to the form
\beq
 am_q \tilde\Gamma_5 (p^2)=A(p^2)+am_q B(p^2)
\label{eq:fitform}
\eeq
where, from the continuum Ward identity, 
one gets for the non-perturbative part: 
\beq
A(p^2)=A_0 \times {[\alpha_s(p^2)]^{7/11}\over a^2p^2} \left[1+22.0\
{\alpha_s(p^2)\over 4\pi}\right]
\label{Ap2}
\eeq
and for the perturbative part: 
 \beq
B(p^2)=B_0 \times [\alpha_s(p^2)]^{4/11} \left[1+8.1\ {\alpha_s(p^2)\over 4\pi}
\right]
\label{Bp2}
\eeq
with $A_0$ and $B_0$ some constants.
These two-loop renormalisation group improved 
corrections are valid for the MOM renormalisation
scheme, in the Landau gauge.
We shall first perform
an extrapolation linear in $m_q$ of the three datasets
to the chiral limit 
for each value of $p^2$.  This fit gives us both $A(p^2)$ and
$B(p^2)$. 
%

%
The lattice data turn out to be quite close to the
continuum theoretical expectations.
Indeed one finds that:\\
$\bullet$ $A(p^2)$ is behaving remarkably close to $1/p^2$ over a large
interval of $p^2$ 
{} The Goldstone contribution appears to be very large:
$a^2p^2 A(p^2) \simeq 0.015$
from the lowest point $a^2p^2=0.16$.
On the other hand, we do not see the log factors expected from
the perturbative calculation. \\
$\bullet$ $B(p^2)$ is found to evolve in good conformity with the
two-loop MOM renormalisation formula quoted above.
We obtain $B_0=1.735$
which provides a very good fit to the data for $p^2$ larger than $2$ \Gev.
\par
%
The Goldstone contribution is felt already at
rather large quark masses and, for physical $u,d$ quarks it is in fact very
large: $A(p^2)$ is larger than the perturbative part $am_q B(p^2)$
even at rather large $p^2$, as shown in Fig.~1. This 
finding for the PS vertex has important consequences.
\begin{figure}[tbh]
\vspace*{-0.3cm}
\epsfxsize=6.5cm \epsfbox{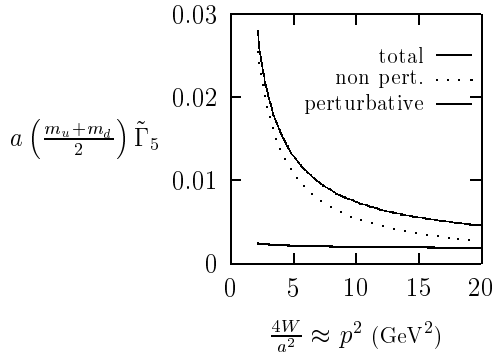}
\vspace*{-0.8cm}
\caption{ The value of $am_q \tilde\Gamma_5$ for light quarks.
}
\vspace*{-0.5cm}
\end{figure}

\section{A large Goldstone pole contribution: consequences on $Z_P^{MOM}$}
%
One can translate the above fit
of $\Gamma_5$  into an expression for $Z_P^{MOM}$,
or rather its inverse which is more directly
physical:
\beq
{1\over Z_P^{MOM}(p^2)}={\Gamma_5(p^2)\over Z_\psi(p^2)}
=
{A_Z(p^2)\over am_q} +B_Z(p^2)
\label{ZPMOM}\eeq
This is to be contrasted with usual fits, which
assume that $Z_P$ is linear in $am_q$.
The fact that the non-perturbative estimate
of the full $Z_P$ differs sizeably from the short distance
$B_{Z}$ already at the measured kappas,
is a signal that
it is not presently possible to work at $p^2$ high enough for the
Goldstone contribution to be negligible. Hence, we must first
subtract it from $Z_P^{-1}$.
\beq
\left[Z_P^{Subtr}(p^2)\right]^{-1}=\left[Z_P^{-1}(p^2)\right]-{A_Z(p^2)\over
am_q}=B_Z(p^2)
\eeq
Numerically at $a^2p^2=1$ we find $Z_P^{Subtr}=0.53$,
which corresponds to the full resummed short-distance
contri\-bu\-tion determined directly from the lattice data.
\section{Consequences on light quark masses}
To get an estimate of the consequences on the 
short distance 
MOM renormalised quark mass given by:
\beq
        am_{AWI}^{Landau}(\mu^2)=\rho {Z_A\over Z_P(\mu^2)}
\eeq
where $\rho$ is a mass parameter,
we use $Z_P^{Subtr}$ and we find in the $\overline{\rm MS}$
scheme :
$a~m_{u,d}^{\overline{MS}} \sim 0.0024$, 
therefore about $4.6$ MeV at $N_F=0$.
Note that these
numbers are only indicative; in view of the many uncertainties
in the subtraction procedure, we do not try to discuss the
other sources of error necessary to give a real determination
of the mass. Our aim is only to underline the necessity of the subtraction
of the Goldstone contribution. \\
This contribution, which is only parasitical 
and has to be subtracted in the calculation of 
$\overline{\rm MS}$ masses,
retains an important
physical meaning in other definitions of
renormalised quark masses.
Let us recall the
Georgi-Politzer renormalisation condition for the quark propagator
\beq
 S_R^{-1}(p,\mu^2)
=i\not p +m_R^{GP}(\mu^2) ~~~~~\mbox{at} ~~~~~p^2=\mu^2
\label{normalisation2}
 \eeq
This defines a renormalised quark mass $m_R^{GP}$ which can be
related to the PS vertex through the axial Ward identity
\beq
m_{AWI}^{Landau}(\mu^2) \Gamma_5^R(q=0,p^2,\mu^2)=
{Tr[S_R^{-1}(p,\mu^2)]\over 4}\label{normalisation}
\eeq
Numerically, using the full $Z_P(p^2)$,
one gets
$a~m_R^{GP}(a^2 p^2=1) \sim 0.018$,
therefore around $34$ \Mev ~at $p=1.9$ \Gev, see Fig.~2.
The magnitude is larger 
than what is expected from the quark condensate and from
the perturbative evaluation of the Wilson coefficient, but 
in agreement with other physical expectations.
\begin{figure}[tbh]
\vspace*{-0.3cm}
\epsfxsize=6.5cm \epsfbox{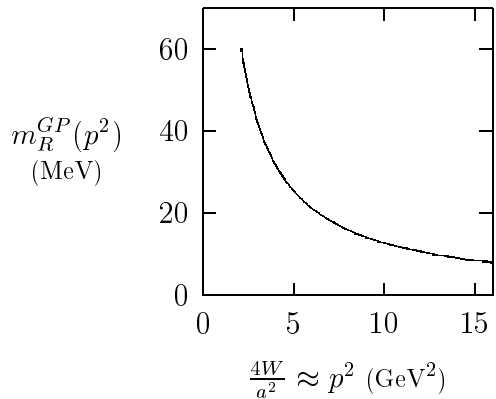}
\vspace*{-0.8cm}
\caption{The value of the dynamical $u, d$ masses.}
\vspace*{-0.5cm}
\end{figure}
We stress that, in contrast to the
standard $\overline{\rm MS}$ current mass, this mass
does not vanish  in the chiral limit, 
%
and it represents
a  dynamically generated mass for light quarks, 
though off-shell, gauge-dependent  and Euclidean. 
This is a well-known 
signal of the spontaneous breakdown of the chiral symmetry.
\section{Relation with the lattice quark propagator analysis}
We have obtained analogous results directly by looking
at the scalar part of the improved quark propagator.
The discussion
of the propagator lattice data involves delicate problems
of improvement, since the 
$O(a^2)$ corrections appear to be dominant
over the standard off-shell $O(``a")$  improvement.
One possible way out is to adopt an improved definition of the
propagator as the solution of a non-linear equation
which has a smoother tree-level $p^2$ dependence, see ref.\cite{prop}.
In Fig.~3 we present our preliminary results,
not yet estrapolated to the chiral limit,
for the ratio between the scalar and the vector part of the 
improved non-linear propagator, obtained by using the data
of the APE collaboration \cite{ape}.
The bare improved propagator is written as
(here and in Fig.~3 the notation $A$ and $B$ do not correspond to the
ones used in the previous sections),
\beq
(S^{imp}(p^2))^{-1}\equiv A^{imp}(p^2)(isin(a\pslash)) + B^{imp}(p^2)
\eeq
hence in this case one has:
$a~m_R^{GP}(p^2)=B^{imp}(p^2)/A^{imp}(p^2)$.
\begin{figure}[tbh]
\vspace*{-0.3cm}
\epsfxsize=6.5cm \epsfbox{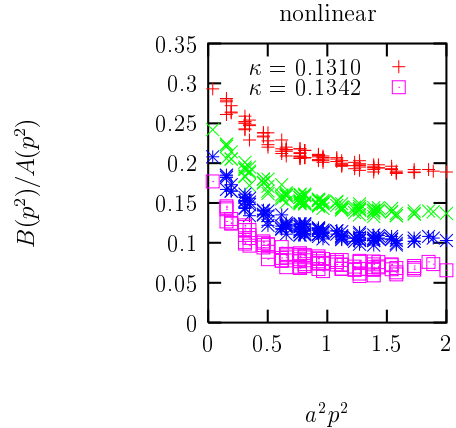}
\vspace*{-0.8cm}
\caption{ The value of $a~m_R^{GP}(p^2)$
as obtained from the non-linear equation for the improved propagator.
}
\vspace*{-0.5cm}
\end{figure}


\end{document}